# A REGULATORY PLACEBO?

## The Systemic Failure of Mandatory GenAI Labeling

**CHEN Jingyi**
https://orcid.org/0009-0005-2175-7680
Faculty of Law, University of Macau, MC 999078, Macau

**BU Chaofan***
https://orcid.org/0009-0006-3384-3962
Research Center for Criminal Justice, Renmin University of China, Beijing 100872, China
Faculty of Law, Waseda University, Tokyo 1698050, Japan

**YAN Shibo***
https://orcid.org/0009-0003-5447-7744
Professional Graduate School, Meiji University, Tokyo 101-8301, Japan

**Corresponding Author: LI Xuesong**
https://orcid.org/0009-0005-6021-6975
Institute of Comparative Law, Meiji University, Tokyo 101-8301, Japan
xuesong.contact@gmail.com

*Authors BU and YAN contributed equally to this work

**Abstract:** We examine the worldwide trend of mandatory labeling of generative artificial intelligence (GenAI) as a reactive, symbolic form of legislation triggered by technological panic and institutional responses. From a technical perspective, this study demonstrates that current mandatory labeling not only creates implementation dilemmas but also risks hindering the evolutionary trajectory of AI technology. We then systematically analyze the three dominant theoretical strands of this regime—the value dilution theory, the information authenticity theory, and the proactive regulation theory—and find that they are products of regulators' cognitive limitations in understanding the logic of modern technology. Not only do such formalistic compliance requirements become a regulatory placebo, but they also obscure the genuine legal demands of the technological era. This challenges the current governance paradigm and suggests a shift from identity-label governance to content governance, with an urgent need to address the complex problems associated with GenAI.

## 1. Introduction: The Global Rush Toward GenAI Labeling

While AI technology has come a long way over the years, it received renewed social attention in 2022 with the sudden emergence of ChatGPT (Nah et al., 2023). Generative AI is not merely a technological breakthrough. It has also triggered widespread global panic and governance anxiety. Nowadays, regulatory authorities in various countries have been quick to react to emerging disruptive technologies, accelerating the evolution of regulation, and labeling is considered a key regulatory measure (Stuurman & Lachaud, 2022).

Recent legislation establishes guidelines to distinguish between AI-generated and human-generated content (El Ali et al., 2024). Service providers are required to add visible or concealed markers on AI-generated content. The main goal is to enable individuals to distinguish between human- and AI-generated content, a task that is becoming increasingly important amid evolving information landscapes (Wittenberg et al., 2024). This strategy aims to safeguard the authenticity of information and preserve human agency. Nevertheless, questions persist regarding the legitimacy and effectiveness of these measures.

The current state of legislation for Gen AI across countries reveals reactive features that stand in contrast to a cautious approach to the law. The rapid pace of the legislative process has led to a fundamental flaw in procedural justice: the absence of a broad social consensus. In the past, key technology laws have been preceded by broad public debate and stakeholder negotiations. However, the current law appears to be a reactive measure aimed at addressing technological disruption. It lacks an adequate legal basis and strong public support. The more serious issue is the rationality of such laws. Is there a good legal basis for this system? Furthermore, technological change is rapid, and it remains to be seen whether a static, identification-based regulatory approach can produce the desired regulatory outcome. These important empirical topics are missing from the current policy-making process, both in their discussion and in forecasting.

Extensive academic research exists on generative AI governance, yet public debates predominantly emphasize improvements to identification systems and technical standards. This article challenges that limited perspective. Rather than dismissing regulation, it advocates for renewed consideration of foundational academic concepts. We must remain vigilant against the regulatory placebo effect, in which legal interventions provide reassurance without addressing underlying challenges.

In this article, the definition of the mandatory labeling system, which is used before the formal argument, is important. Not all AI transparency regulations are included in the discussion. This targeted source labeling/traceability is acknowledged to be auxiliary in high-risk situations such as deepfakes, synthetic pornography, political manipulation, financial and medical advice, and judicial evidence. This article aims to challenge a system design with the following three features. Firstly, it only applies to the use of generative AI. Secondly, if there is no label, it is a legal violation, and there is no distinction between technical and substantive violations. Thirdly, the system has ambitious goals: to protect the authenticity of information, restore public trust, and ensure accountability, rather than directly addressing information risks. Ultimately, the central question is whether AI labeling should be universally mandated and whether non-compliance should be treated as a systemic governance issue.

## 2. Global Regulatory Resonance: The Illusion of Consensus in GenAI Labeling

The technological disruption introduced by ChatGPT has triggered a global regulatory response, leading to competitive initiatives among regulatory authorities. The legal norms developed during this period represent efforts to delineate the boundaries between artificial and intelligent systems in a rapidly changing algorithmic landscape. While the intensity and character of this regulatory surge differ across jurisdictions, a collective sense of urgency is evident.

### 2.1 Japan: The Self-Regulatory Approach

Japan has no mandatory AI labeling regulation. The central policy, the AI Guidelines for Enterprises, is coordinated by the Cabinet Office's AI Strategy Conference and jointly issued by the Ministry of Economy, Trade and Industry (METI), the Ministry of Internal Affairs and Communications (MIC), and additional government agencies (Ministry of Economy, Trade and Industry, 2024). The policy does not establish binding legislation for AI labeling; instead, it encourages companies to implement a risk-based approach to autonomous governance. To mitigate the risks of misinformation and intellectual property infringement associated with Gen AI, the guidelines advise AI developers and providers to incorporate digital watermarks or explicitly label content as AI-generated to safeguard users' right to information. This approach demonstrates Japan's effort to balance technological innovation with risk management and aligns with international initiatives, such as the G7 Hiroshima AI process, to enhance transparency in generative AI (KeguanJP Editorial Department, 2024).

### 2.2 EU: Comprehensive Transparency Mandates

The European Union, as a leading exporter of global digital regulatory frameworks, implemented new regulations in 2024 that demonstrate its ambitious institutional approach. Article 50(2) of the Act establishes universal transparency standards for Gen AI service providers and mandates that their outputs be labeled in a machine-readable format. This requirement extends beyond mere information disclosure, aiming to translate authenticity into a physical parameter that systems can detect in accordance with the underlying standard. While the main text of the bill does not specify all technical details, Recital 133 presents an idealized regulatory blueprint for watermarking, metadata, and fingerprinting technologies. This emphasis on technological robustness and interoperability illustrates the EU's strategy to address increasing digital governance concerns through standardized technical measures.

### 2.3 US: Fragmented Governance and Local Experimentation

In contrast to the European Union's holistic regulatory framework, the United States' legislative landscape illustrates a dynamic interplay of negotiation under reactive legislation. At the federal level, regulatory approaches continue to rely predominantly on voluntary commitments and soft-law guidelines, as outlined in the 2023 presidential executive order. This approach lacks legal enforceability and is widely regarded as a political reaction intended to reassure the public. Nevertheless, assertive local interventions have altered this equilibrium. For example, California's Transparency in AI Act (CAITA) signals a significant transition in the United States from reliance on voluntary compacts to the implementation of enforceable state legislation (Andrew Simmerman, 2026).

CAITA demonstrates a deliberate approach to selective regulation. By establishing a threshold of more than 1 million visitors or users per month in California, the regulation specifically targets large generative AI service providers (California State Legislature, 2025, sec. 22757.1(b)). The regulation applies exclusively to images, video, and audio, explicitly excluding plain text output. Exceptions are made for entertainment content, including non-user-generated television, movies, streaming, and video games. This approach, which prioritizes oversight of major providers while exempting smaller entities, reduces compliance costs for start-ups and textual tools. It also reinforces the role of human intervention in taxonomic frameworks. If the primary risks are inherent to algorithmic generation, this dual exemption based on provider size and content type may reflect a regulatory stance intended to address a perceived public concern, rather than a differentiated approach grounded in risk assessment (Omaar, 2024).

### 2.4 China: End-to-End Strict Compliance

China demonstrates leading administrative efficiency and comprehensive reach in constructing its identification system, resulting in a rigorous framework of departmental regulations and mandatory national standards. The "Measures for the Identification of Artificial Intelligence-Generated Synthetic Content," which takes effect in September 2025, formalizes a governance system that shifts from guidance to robust control (Cyberspace Administration of China, 2025). The system is based on mandatory implicit identification through metadata to enable traceability, and on explicit identification in specific cases to avoid public confusion. Its goal is to provide for identification at any time, anywhere in the information domain. Service providers must ensure that attribute information, the provider's name or code, and the content identifier are embedded into the metadata of the synthetic content, in accordance with Article 16 of the Provisions on the Administration of Deep Synthesis of Internet Information Services. The regulation is based on the idea of technical identifiability. China has created a highly closed-loop responsibility system: labeling is required at both the content generation stage and the dissemination and distribution platforms. The impact of this legislation, however, could be undermined by the fluidity of generative technology and the challenges posed by its decentralized, intangible evolution, casting doubt on the proposed full-chain regulation.

### 2.5 Korea: Dual-Track Labeling Legislation

Korea has taken a different route than Japan, China, and the European Union, passing broad AI regulations with two requirements: hierarchical supervision and universal disclosure. The "AI Basic Law," which came into effect on 22 January 2026, is the second comprehensive law on AI worldwide, after the European Union's Artificial Intelligence Act (Ministry of Science and ICT, 2024). The Act uses a risk classification system and requires prior notice, manual oversight, and an impact assessment for high-impact AI use across industries, including healthcare, energy, transportation, credit approval, and recruitment. Further, non-compliance with universal notification and labeling obligations for AI-generated content is punishable by a fine of up to 30 million. An implementation grace period of at least 1 year has been set to facilitate a smooth transition (Beijing DHH Law Firm, 2026 February 3).

From a critical perspective, Korea's dual-track regulatory structure offers significant value by providing two distinct approaches within a single legal framework. For high-impact AI, the legislation assigns obligations to specific high-risk scenarios, such as medical, financial, and administrative contexts, thereby implementing a differentiated, risk-based disclosure system with recognized ancillary value. In the case of generative AI, by contrast, legislation has established a universal requirement to inform and label all products and services based on AI-generated output. This is an example of the 'labeled' approach to risk assessment through source identification, a practice that has been criticized. A question now arises: If the law imposes the same source obligations on a full-compliant market analysis conducted by an AI and on a deepfake video designed to manipulate the market for political purposes, does the regulatory system still send risk signals? The design of the dual-track in Korea seems logically correct. However, the universal notification for generative AI is still subject to the placebo effect of the formal notification, with no differentiation.

### 2.6 The Underlying Convergence Beneath the Global AI Regulatory Boom

**Table 1** and the analysis above show that the five jurisdictions do not have consistent mandatory labeling requirements. Existing international standards fall into five different models. In Japan, soft law and business self-regulation are the two instruments. The EU incorporates machine-readable marks in its transparency requirements and specifies exceptions. China uses a holistic strategy that includes both explicit and implicit indicators. California's regulations target big generative AI companies. Korea has established a dual-track approach to the use of high-impact and generative AI under the AI Basic Act, with a grace period of at least 1 year before implementation.

| Jurisdiction | Regulatory Approach | Stringency Level | Targeted Entities | Content Types | Primary Objectives | Exemption Mechanisms |
|---|---|---|---|---|---|---|
| **Japan** | Soft law, Self-regulation | Low | Enterprises / Business entities | General AI use | Risk governance, Balancing innovation | Voluntary |
| **EU** | Risk governance + Transparency obligations | Medium-High | Providers / Deployers | Text, image, audio, video (with exceptions) | Anti-deception, Information integrity | Exceptions for standard editing, human review, etc. |
| **US Cal.** | State law, Platform / Service provider obligations | Medium | Large GenAI providers | Primarily image, audio, and video | Provenance authentication, Detection tools | Exclusion of certain entertainment content |
| **CN** | Departmental regulations + Visible & invisible labeling | High | Service providers, Dissemination platforms, Users | Text, image, audio, video, virtual scenes | Content governance, Platform liability, Traceability | No explicit labeling under agreement, but logs are retained |
| **Korea** | Comprehensive legislation + Dual-track obligations | Medium-High | High-impact AI entities, GenAI service providers | GenAI output content | Transparency obligations, Anti-deception, Building trust foundation | Artistic creation content |

**Table 1:** The Divergence of Global GenAI Labeling Mandates

The five models differ significantly in the stringency of their requirements, the entities subject to compliance, the types of content regulated, and the nature of their exceptions. The differences illustrate how, while there is a global trend towards increased regulation, different jurisdictions focus on different risks. In the EU, for instance, specific exceptions are provided, while in California, boundaries are drawn around media. In Korea, a distinction is made between high-impact and generative artificial intelligence. This analysis examines whether, in the absence of tailored approaches and under universal labeling mandates, such measures effectively safeguard authenticity or merely serve as a response to political pressure. As illustrated in **Figure 1**, this reactive legislative resonance risks trapping global governance in a vicious cycle of a REGULATORY PLACEBO.

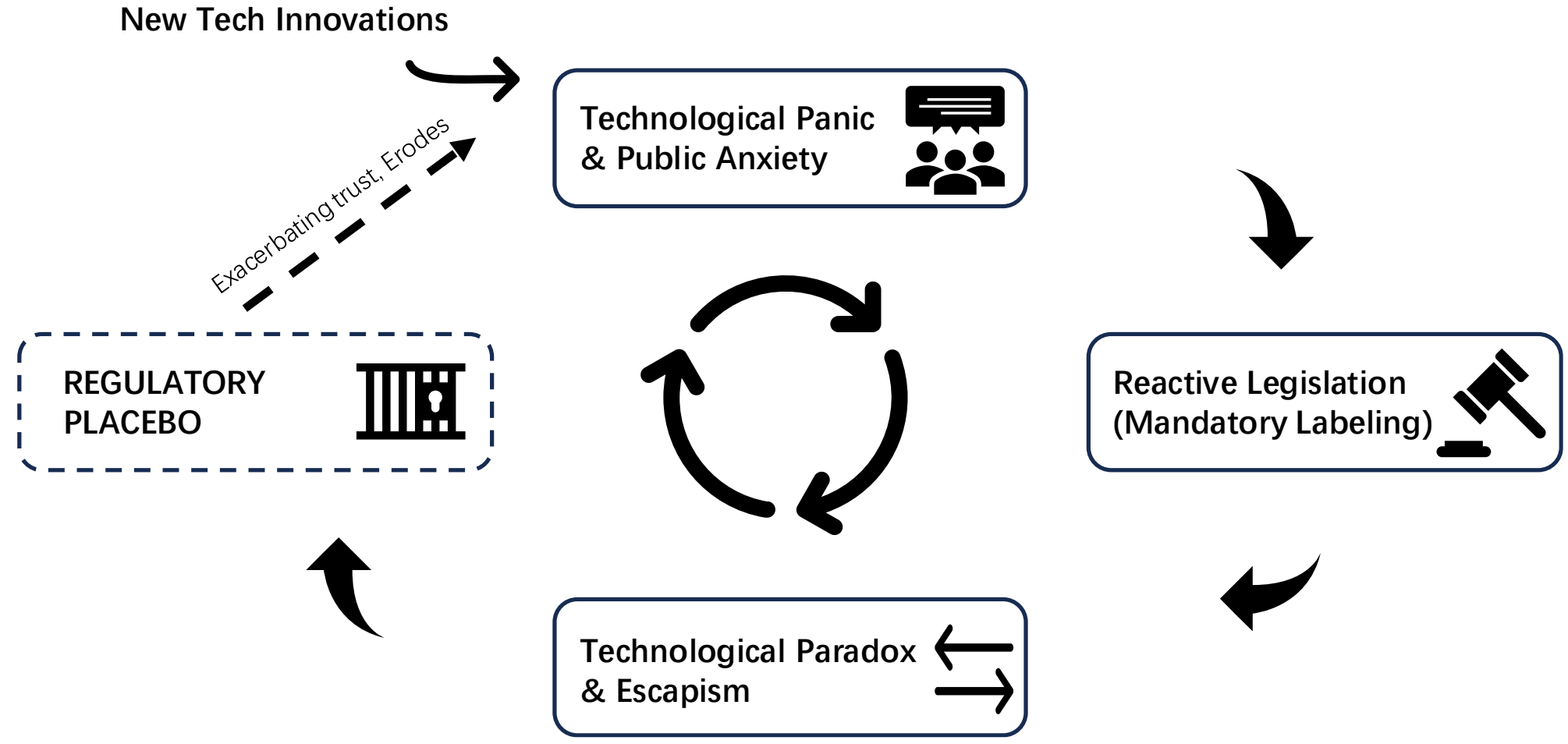

**Figure 1** The Vicious Cycle: A Self-Reinforcing Regulatory Loop

## 3. Substantive Unfeasibility: Implementation Crises and Technological Paradoxes

This chapter addresses whether the norms can be implemented, while Chapter 2 considers reactive administration at the normative level. The question of “should not” is reserved for Chapter 4. Before diving into a formal discussion, it is important first to understand what is meant by content generated by generative AI. AI content raises questions such as whether AI-written and human-edited content is eligible, whether AI translations are followed by human editing, and whether content is deemed AI-generated if AI provides several edits and the human chooses some (Simard, 2024). Furthermore, the extent of AI use, for example, whether it is used to edit grammar, is unclear. At present, there are no clear limits in national legislation on this basic semantic question. The overarching concept in the EU's AI Act is "generated or manipulated by AI." China’s labeling requirements apply to content that is created, produced, or significantly rewritten. It is defined in California’s SB 942 as generated or modified by GenAI systems.

While these definitions seem straightforward in the text, there is nonetheless a high level of human-machine collaborative writing, and the regulatory authorities will ultimately be tasked with making the decision. Even if legislators try to put in place a firm limit, e.g., requiring identification for AI contributions exceeding 50%, this threshold cannot be precisely defined. There is currently no reliable technical solution for determining the relative proportions of human and machine contributions to a text. This means that if there is no clear definition of what is regulated, there is no basis for enforcement or for the predictability of a legal obligation.

Even if the definitional dilemma is resolved and content can be identified as AI-generated and labeled in accordance with legal requirements, two fundamental questions persist: whether national law can effectively regulate the intended object (Section 3.1) and whether current technological approaches can meet the legal expectations for such regulation (Section 3.2).

### 3.1 The Regulatory Illusion: Jurisdictional Escapism and the Void of Punishability

#### 3.1.1 How GenAI Evades Physical Boundaries

A central question in assessing the effectiveness of mandatory labeling for generative AI is: What can regulators effectively monitor? Illegal activities are not always preventable by traditional regulation, but regulated activities tend to have fixed physical carriers, identifiable transactions, or accountable entities. Generative AI, on the other hand, lives in a dematerialized world where its production, reproduction, and dissemination lack tangible roots. Consequently, traditional regulatory mechanisms based on tangible objects, fixed boundaries, and identifiable actors may prove to be empty gestures.

##### 3.1.1.1 The Dual Failure of Physical Control and Platform Control

The regulation of physical goods depends on two overlapping structural conditions. First, goods must pass through ports and borders, where authorities can inspect and detain shipments at designated checkpoints (Leuprecht, 2021). Second, e-commerce platforms serve as concentrated distribution nodes. Although the circulation of goods cannot be entirely blocked, authorities can leverage platforms such as Amazon as a secondary checkpoint by mandating prior review of merchant qualifications and product labeling, and by enforcing the removal or banning of products that violate regulations.

The deployment and utilization of generative artificial intelligence services undermine the previously established conditions. The generation, transmission, and consumption of content depend on intangible data flows that bypass physical channels, rendering traditional customs-based controls ineffective. Consequently, regulators have shifted their focus to controlling critical points within information infrastructure by mandating that gatekeepers, including social platforms, app stores, content delivery networks, and cloud service providers, assume responsibility for compliance reviews and attempt to block unidentified content during distribution.

However, this path has structural limitations. First, generated content does not necessarily rely on large platforms for dissemination. Small websites, decentralized social networks, peer-to-peer communication, and even e-mail can carry unidentified content. These long-tail nodes are naturally free from gatekeeper supervision. Second, the watermark can be removed, tampered with, or scrubbed through screenshots, text extraction, and other means. Even if the source fulfills its labeling obligations, it remains difficult to prevent the content from re-entering the public space in an unidentified state (Chen, 2024).

Third, as hardware costs decrease and the open-source community advances, individuals and small- to medium-sized organizations are increasingly able to run large language models with advanced generative capabilities on local devices, eliminating the need for centralized servers or platform compliance procedures. Locally generated content can subsequently be disseminated in the public sphere through USB drives, local area networks, or reconnected social media, and is indistinguishable from other content (Zheng, 2025).

A regulatory framework that depends exclusively on centralized service providers is consistently inadequate when addressing decentralized technology applications.

#### 3.1.1.2 Jurisdictional Vacuums and Self-Imposed Constraints

Even if the regulatory approach is technically feasible, there is an inherent tension between the territorial scope of jurisdiction and the Internet's borderless nature.

At present, Japan’s Fugaku-LLM, the United States' models, including GPT, Gemini, and Grok, and the Chinese DeepSeek are all proceeding apace. The concept of territorial jurisdiction in public international law means that a state’s laws apply only within its territory (Abdelaziz, 2025). Without a binding international agreement on AI governance, national regulations face two problems. First, domestic legislation often struggles to regulate foreign service providers effectively. These entities can inject unlabeled content into the domestic market via cross-border data flows without establishing a local corporate presence. Second, extraterritorial enforcement of labeling requirements through long-arm jurisdiction may lead to legal conflicts and governance issues with other states (Sheng & Hong, 2023).

Consequently, strict local labeling requirements constrain domestic industries’ flexibility in complying but fail to prevent the ongoing influx of unidentified content from external sources. This dynamic effectively establishes a regulatory framework characterized by unilateral self-restraint.

### 3.1.2 Punishability Basis

Despite the two dilemmas, it is possible to posit an ideal scenario in which law enforcement has complete knowledge and can accurately identify every actor who has failed to fulfill the identification obligation. Under such conditions, the feasibility of perfect law enforcement for AI identification warrants examination.

In criminal law theory and administrative punishment jurisprudence, the core doctrine of punishability is typically grounded in the frameworks of social contract theory[1], infringement of legal interests[2], or the abstract danger structure.[3] The legitimacy of public power exercised by regulatory authorities to impose sanctions on individuals should be predicated on conduct that violates the common social contract or causes substantial or highly foreseeable harm to a specific legal interest.[4] Regarding generative AI identification obligations, if AI-generated text or images are neutral or accurate in content and do not result in significant harm to social order or the rights of others, the sole deficiency lies in the absence of an AI-generated label.

From the perspective of regulators, the obligation of identification serves to uphold procedural legal interests, the recipients' right to know, and the maintenance of societal trust. To substantiate this proposition, at least two conditions must be satisfied.

First, legislation must reasonably justify the extent to which the public substantially relies on knowing whether information is AI-generated. If this reliance is not supported by empirical social experience, treating all unlabeled content as an infringement on the right to know becomes a flawed assumption.

Second, even if an abstract danger is acknowledged, one must further evaluate whether the sanctions align with the principle of proportionality—particularly when the absence of a label is not compounded by substantive offenses such as fraud, defamation, counterfeiting, or manipulation of public opinion.

In the absence of a clear legal basis and supporting empirical evidence, the labeling obligation risks degenerating into a mere box-ticking exercise. Consequently, actors are constantly forced to attach labels to prove their innocence. However, this burden of proof does little to improve the information ecosystem or meaningfully enhance public trust. Eventually, the labeling system could transform from a governance tool designed to increase transparency and accountability into a formalistic burden lacking a solid foundation of legitimacy. Should these practical dilemmas of technological uncontrollability and selective enforcement materialize, the labeling system will not only fail to mitigate the actual risks of GenAI. However, it may also undermine the consensus regarding the jurisprudential basis for punishment and proportionality.

### 3.2 Technical Limitations and Logical Paradoxes: The *Ought Implies Can* Dilemma

If the boundaries of legislative jurisdiction limit the scope of regulation for the identification system, the logical paradoxes of technology may undermine the system’s epistemological foundation. Notably, technical feasibility does not necessarily equate to theoretical legitimacy or the effectiveness of institutional design.

While Section 3.1 addresses obstacles encountered by the identification system at the jurisdictional and enforcement levels, this section examines a more fundamental failure. Even if all jurisdictional barriers were eliminated and the law could precisely monitor all content, the inherent logic of identification technology makes it hard to achieve the governance objectives.

---

[1] Social contract theory posits that an act is punishable only when it violates the fundamental norms of the social contract, and such punishment is required to preserve societal order.

[2] The theory of infringement of legal interests holds that if an act violates moral standards but does not infringe upon legal rights, it is not subject to criminal punishment.

[3] Abstract danger denotes a legal presumption of risk, whereby the commission of an act is deemed punishable without the necessity of demonstrating actual harm.

[4] pecific legal interests encompass personality rights, property rights, public order, safety, and trust.

Recent advancements in computer science have produced AI detection technologies, including frequency-domain invisible watermarks, metadata embedding, vocabulary-statistics-based analysis, and the distribution of specific high-frequency words (Kirchenbauer et al., 2023). Technically, distinguishing between human- and machine-generated outputs is a feasible engineering challenge. However, the primary concern lies in the intended purpose of implementing these technologies. If the distinction is pursued solely for identifiability, rather than for risk governance, liability tracing, or rights protection, this technical approach lacks substantive value. Furthermore, regulatory measures should avoid restricting the capabilities of generative AI within technological governance. Otherwise, the system may experience internal limitations rather than external opposition.

This section examines five progressive approaches, each converging on the conclusion that no technical method exists to identify generative AI. Technical advancements alone cannot resolve any single approach. The failure is multi-layered, structural, and mutually reinforcing. Even if one problem layer is addressed, the remaining layers persist.

### 3.2.1 The Pareto Trade-off Between Robustness and Generation Quality: The Intrinsic Paradox of Identification Technology

Large language models face a fundamental conflict between the need for mandatory labeling and the goal of achieving human-like language proficiency (Grinbaum & Adomaitis, 2022). Generative AI seeks to replicate human natural language patterns, aiming to minimize the disparity between machine-generated and human-authored content to the point that the two become nearly indistinguishable. In contrast, current text detection and digital watermarking methods deliberately introduce statistical biases or distinctive machine-specific markers[5] into generated outputs. These markers can be detected for identification purposes and are often designed to be nearly imperceptible to human readers. This approach, however, introduces a Pareto trade-off between concealment and robustness.[6] Enhancing the reliability of mandatory identification technologies typically diminishes the quality of AI-generated content and increases computational demands. Conversely, reducing the strength of identification measures to maintain naturalness leaves the system susceptible to ineffectiveness, such as word substitutions or punctuation changes (Liu et al., 2024).

This leads to an antinomy at both the technical and the normative levels—a Kantian paradox in which two mutually exclusive principles appear equally logically justified. From a technological perspective, as AI becomes more sophisticated, its output increasingly mirrors the natural distribution of human text, making it nearly impossible to distinguish machine-generated content from human writing. By contrast, in regulatory compliance, models must be traceable in their language output so they are subject to oversight (Liu et al., 2024). Consequently, current mandatory labeling practices effectively attempt to de-anthropomorphize these technologies, artificially counteracting their natural evolutionary tendency to become more human-like.

This approach compels models to sacrifice intelligence for regulatory compliance, creating a dynamic that may ultimately undermine technical advancement. If legal frameworks mandate that AI-generated works consistently exhibit statistically identifiable non-human characteristics, this requirement directly conflicts with training AI models to produce natural language. Such a mandate imposes a permanent limitation on the linguistic evolution of these models, generating structural tension between regulatory transparency and linguistic competence. While this may not entirely obstruct the development of artificial intelligence, as models could potentially circumvent watermarks through advanced training, it substantially increases engineering costs. It distorts the natural trajectory of model evolution.

[5] Machine-specific markers refers to the software and hardware information collected from a device for identification purposes.

[6] Robustness is an important feature of an AI system, algorithm, or model that enables it to maintain stable performance in the face of external environmental interference or data anomalies.

#### 3.2.2 Data Contamination: Systematic Distortion of Model Evolution Paths by Identification Technology

A significant concern is the risk of data contamination, which varies in severity across different models. The iterative development and upgrading of generative models depend on the continuous extraction of new corpora and parameter updates from open networks. For images and videos, watermarks are often embedded with pixel-level frequency robustness to perturbations and good statistical persistence across large-scale retraining datasets. As hidden watermarks from various vendors increasingly dominate internet images, they create artificially distorted probability distributions. Inevitably, these distortions will be ingested by next-generation visual models, which will process them as normalized signals during subsequent training cycles. Watermarking is generally implemented in the text domain using a token-level logit bias technique, which is not resistant to basic text processing such as rewriting, translation, secondary pasting, etc. This means that contamination in textual training data will be systematic and statistical, rather than manifest distortions. Moreover, if a large amount of generated content continues to be added to training corpora, regenerated by the next version of the model, recycled through the network, and retrained, the bias of the identification technologies will continue to affect the next version of the model's perception of natural language distributions. The magnitude of this distortion will depend on the effectiveness of the training data-cleaning mechanisms, but its direction will be the same as the closed loop.

Over time, models trained on natural language or image patterns may develop systematic biases. Rather than learning from unaltered natural signals, these models are exposed to data shaped by policy objectives and regulatory requirements. This process can reduce the ability to generalize and robustness, while also posing significant institutional challenges to the development of artificial general intelligence. Despite its importance, this issue is seldom addressed in current policy discussions, although it has enduring and potentially irreversible effects on technological progress and the broader discourse ecosystem.

#### 3.2.3 False Positives and Structural Discrimination: Who Is Paying the Price for Algorithmic Misjudgments?

Text detection methods that rely on statistical patterns are at significant risk of misidentification in practical applications, particularly within professional standardized writing. Generative AI models, trained on extensive human corpora, produce outputs that restructure and simulate established human language patterns (Zyda, 2024). In formal professional domains such as legal documents, official communications, and academic works, authors adhere to strict conventions. Over time, these domains have developed highly formalized and formulaic conventions of expression, resulting in writing that is inherently boilerplate. When models are trained on large volumes of such professional texts, they internalize and frequently reproduce these standardized expressions. This situation creates a paradox: the more rigorously legal professionals or civil servants adhere to professional standards, the more their language aligns with model-generated outputs, thereby increasing the likelihood of being misclassified as AI-generated. It is unreasonable to expect legal professionals to abandon longstanding writing conventions to circumvent the limitations of current AI detection. In this context, detection algorithms not only fail to differentiate effectively but also penalize those who adhere strictly to norms, a consequence that is unacceptable in judicial and administrative contexts.

Comparable challenges are observed among language learners and cross-language writing groups. The implementation of mandatory labeling significantly undermines equality for second language learners within the writing community. Individuals acquiring a second language often write or translate using a word-by-word processing pattern. This approach, characterized by rigid word order, fixed-vocabulary collocations, and an unnatural sense of language, closely parallels early machine translation systems and current large language models in generating low-resource outputs. Such parallels reflect broader principles of human language cognition. In translation contexts, if they regularize discrimination on subjective criteria such as rigidity, template, or unnaturalness to classify text as AI-generated, they risk institutionalizing discrimination against non-native speakers and other vulnerable groups in educational settings. In competitive environments, including examinations and study abroad applications, the institutionalization of this discrimination through official test reports further exacerbates inequality.

In this context, directly converting test results characterized by statistical uncertainty into evidence for judicial or administrative proceedings lacks a rigorous evidentiary foundation. The admissibility of evidence requires fundamental criteria, including repeatability, explainability, and controllable error rates. Current artificial intelligence detection tools exhibit significant deficiencies in these respects. First, most detection models are closed-source, and their algorithmic mechanisms lack transparency, making it challenging for parties and judges to assess statistical assumptions and error margins. Second, test results are typically expressed as probabilities, which complicates their translation into the high-probability standard required for legal judgments. Third, the consequences of false positives in detection algorithms are asymmetrical, potentially harming legal professionals and researchers who comply with professional standards or unfairly labeling linguistically vulnerable groups as AI cheaters. Employing such results as the primary basis for sanctions, degree revocation, or denial of examination outcomes presents a false appearance of technological neutrality and conceals emerging structural injustices. Accordingly, from the standpoint of evidence law and procedural justice, these testing tools should be used with considerable caution and restraint.

#### 3.2.4 Image and Video Domain: The Illusion of Watermark Effectiveness and Vulnerability to Countermeasures

Relying on invisible watermarks to regulate AI-generated images and video results in both diminished communication effectiveness and limited technical robustness. For a labeling system to function effectively in public communication, audiences must receive immediate notifications at the point of content exposure, such as a prominent prompt displayed on the browser or platform interface. However, major manufacturers have developed different watermarking solutions without unified decoding standards or cross-platform protocols. For example, watermarking systems from Google, OpenAI, and Meta are incompatible with one another (Zyda, 2024). Unless all vendors' decoders are pre-installed at the operating system or browser level and unified access is provided across applications, these hidden identifiers remain undetectable to most users at points of information dissemination. Recognition is typically possible only in controlled laboratory environments using complex software stacks. Consequently, in the rapid and fragmented landscape of social media, it is nearly impossible to prevent the spread of disinformation. Invisible watermarks function primarily as post-event forensic evidence rather than as real-time cognitive interventions. Without a clear institutional design, governance risks devolve into a situation in which technological solutions are substituted for effectiveness.

From the perspective of an information adversary, detectability often implies both reversibility and vulnerability. Artificial intelligence can employ specific algorithms to embed noisy watermarks in images or videos. Counter-adversarial techniques, including pixel disturbance, filter overlay, denoising algorithms, resampling, screenshots, and secondary compression, can remove or fabricate watermarks at minimal cost (Deshpande & Kanti, 2025). Within this offense-defense dynamic, each additional technical requirement imposed by regulators can be countered by evaders with a simple batch-processing step[7] in their toolchain, creating a significant cost asymmetry. More critically, once a watermark is removed or maliciously altered, the content can easily be fraudulently claimed as human-authored. There is also a risk of framing, in which AI watermarks are forcibly embedded in others' content to indicate that it is AI-generated falsely. Consequently, the distinction system under development is unstable. The recognition of content as AI-generated depends not on its actual creation process, but on the survival or manipulation of a fragile watermark by third parties. Such institutional foundations are unlikely to provide the stability and predictability necessary for judicial and administrative decision-making.

[7] Batch processing tasks are programs that execute a series of commands on a computer without human intervention.

### 3.2.5 The Ideology of Technical Regulatory Pathways: Normative Reflections on Carbon-Based Centrism

The logic that enforces a strict distinction between AI-generated works reflects a carbon-centrism that lacks ethical and ideological self-reflection. Carbon-centrism presumes that all life must be chemically based on carbon, thereby excluding or diminishing the legitimacy of other potential life forms (Susen, 2022). Regulatory demands for special labeling of AI-generated content are often justified by claims that such works are uncertain or more likely to be false or harmful than those created by humans. This rationale, however, overlooks the significant qualitative differences within human-generated content. Humans are also capable of mass-producing low-quality, false, or malicious information, yet contemporary legal systems do not mandate risk-based labeling for all human authors (Chen & Shu, 2024). Current norms do not impose special attribution requirements based solely on the creator's human identity. Requiring content labeling exclusively for outputs produced by silicon-based computing, while disregarding inferiority and malice in human creation, constitutes a double standard and undermines the principle of technological neutrality.

The industry has developed a technical approach distinct from mandatory watermarking, exemplified by the content source signature mechanism defined in the C2PA standard. This mechanism documents the creation, modification, and publication history of content through a cryptographic signature chain. In contrast to statistical watermarks, it possesses several fundamentally different technical characteristics. First, it does not depend on identifiable non-human traces and therefore does not cause the problem described in Section 3.2.1. Second, its effectiveness relies on verifiable sources, which are more detectable than statistical ones, thereby avoiding the false-positive issues outlined in Section 3.2.3. Third, it is voluntary, auditable, and signed proactively by the publisher, thereby preventing it from encountering the territorial jurisdiction and decentralization challenges discussed in Section 3.1. This comparison demonstrates that while mandatory labeling legislation continues to seek to incorporate statistical watermarks into legal obligations, the technology market is advancing toward solutions that adhere to technological logic and fulfill the intended function of disclosure obligations. Continued legislative insistence on the current approach risks impeding superior technological advancements.

The mandatory labeling system imposes highly asymmetric compliance costs on entities. Large service providers such as OpenAI, Google, and Anthropic possess substantial engineering resources to develop watermarking infrastructure, detection tools, and cross-platform compatibility solutions (Reuters, 2023). In contrast, open source communities, small- and medium-sized developers, and individual researchers find it hard to bear the compliance costs. As a result, these smaller entities are more likely to abandon or avoid publication (Rijsbosch et al., 2025). Consequently, mandatory labeling may protect large companies while marginalizing smaller participants. This dynamic incentivizes increased market concentration rather than advancing identification technology. Furthermore, while the system aims to mitigate risks in the information ecosystem through universal labeling, it ultimately suppresses the ecosystem's diversity.

Therefore, from the perspectives of regulation and innovation promotion, a more effective approach is to shift the regulatory focus from the content source to the content's risk. When information poses significant harm due to falsehood, incitement, or manipulation, the same risk assessment and accountability mechanisms should be applied regardless of whether the content is generated by humans or by artificial intelligence. Content should not be pre-classified as low-level information solely on the basis of its origin. Prioritizing risk and effectiveness as the primary classification criteria enables a more rational discourse for distinguishing between human- and machine-generated products, thereby facilitating technical and normative alignment.

## 4. Deficit of Normative Legitimacy: Value Dilution, Information Authenticity, and Proactive Regulation

Chapter 3 demonstrates the structural failure of the mandatory labeling system on the level of enforceability and technical logic. However, this critique is restricted to instrumental rationality. A potential counterargument suggests that, although current technology is imperfect, the labeling system could be feasible and viable as technology advances. Chapter 4 explores a more fundamental issue. Even if there were a perfect technical solution that resolved all enforcement challenges, its normative legitimacy would remain questionable. In this chapter, we examine the three main arguments, which are "value dilution," "information truth," and" positive regulation." It is argued that each exhibits fundamental logical flaws in its legal basis, rather than technical limitations.

### 4.1 Refutation of Value Dilution Theory: Wrong presuppositions based on the labor theory of value

One argument supporting mandatory identification is the value-dilution theory. This perspective assumes that generative AI can produce vast amounts of text or images in a short period with minimal input, compared with the time, effort, and intellectual input required for human creation. The resulting low marginal cost has generated considerable concern, particularly when viewed through the lens of the labor theory of value. Content created without human labor is seen as low-cost and potentially disruptive. Consequently, there is a need for mandatory identification to distinguish such content from high-cost human creations and to prevent inferior content from displacing superior works in the market (Raj et al., 2026).

However, this regulatory logic contains significant fallacies in both economics and jurisprudence. It incorrectly equates low marginal costs for users with an absence of technical costs. Supporters often claim that AI-generated products are unfair because people pay little or nothing for tools like ChatGPT. However, when we consider the entire technology cycle, generative AI is not free. Pre-training large models, fine-tuning with supervision, and modification require significant investments in computing power, hardware, and high-quality data. The efficiency of AI output results not from circumventing the law of value, but from shifting the cost structure from high marginal costs to significant fixed capital investment and low marginal costs. If the marginal cost of final delivery is lower than that of human labor, labeling the output as problematic and requiring warnings is analogous to insisting that industrial assembly-line products be marked as non-handmade to warn consumers. Such reasoning contradicts the fundamental principle of technological progress.

A standard based on production costs or production efforts is inappropriate. Some examples of this include treating poor human content as exempt from labeling because it took a long time to create, and considering good AI-generated content as less valuable because it was produced quickly. This approach is more about protecting both human resources and the regulatory framework rather than merit. Legally, no matter how creative the work is, it should be protected by intellectual property and personality rights. The use of value dilution as a criterion for labeling is a step backward.

In general, we contend that the idea of value dilution concerns market value rather than providing a robust legal foundation. In certain situations, for example, when consumers buy art, custom designs, educational evaluations, or attribution services, they may be interested in the creative process. The degree of human involvement in the production of a work may affect consumers' judgments. Hence, information on the production process is relevant. However, this relevance is largely limited to process disclosures, anti-fraud rules, or contractual integrity requirements for specific transactions. It does not imply that all AI-generated content needs to be universally identified because of low marginal costs. Legal rules should target misrepresentation and transactional deception, not inefficient content production, as a problem to be addressed.

**4.2 Refutation of Information Truth Theory: The Misalignment of Ontological Identity and Truth Attributes**

A second compelling argument supporting the obligation to label is the "truth of information theory." This theory asserts that AI-generated content carries a risk of hallucination and may produce false information. Given the public's right to accurate information, such content should be clearly labeled (Ghaffary, 2023). Although this perspective appears to safeguard the right to know, it misunderstands the difference between the ontological origin of information and its authenticity.

When evaluating the veracity of information, its origin should not be the sole criterion. While AI outputs can be manipulated, humans also frequently generate misleading content due to malicious intent, flawed memory, or cognitive limitations. Legislation aimed at countering disinformation must be based on a screening system for false information, not a presumption against information from AI sources. This means that there is a logical contradiction. If AI-generated text is factually accurate, it is unjustified to presume it will engender public distrust simply because of its algorithmic origin. However, if a human-written article contains false information, it should not be human-made and should not be warnable. A law that targets the source of the information rather than the information itself is an example of the genetic fallacy.[8] It is the legal term for determining whether a viewpoint, claim, or issue is based on its source, background, or history, rather than its content or evidence.

Some scholars contend that, to protect users' right to know, all technologically synthesized content constitutes an alteration of physical reality and should therefore be explicitly identified. If this strict "one-drop of blood principle"[9] is adopted, whereby any content containing artificially generated elements is deemed untrue, the regulatory logic may become excessive. What are called established practices include green screen synthesis for film, post-production photography grading, news imagery editing/composition, and even stage magic. If an AI-created product needs to be labeled, then these old-fashioned technological practices need to be regulated as well. Otherwise, any such attempt is liable to lead to a double standard for new technologies. It is difficult to gain real-world experience because all information is communicated through different media. Even if an AI label could be used to draw such lines, it would be impractical and amount to a simplistic notion of truth in philosophy.

AI-generated sources do not inherently present a risk. However, compulsory labeling may have legal significance in specific contexts but not in all situations. The necessity of compulsory disclosure depends on three key considerations:

[8] Genetic fallacy, also called origin error or origin conclusion, is an informal logical fallacy. It involves judging a viewpoint, claim, or issue based on its source, background, or history rather than its content and evidence.

[9] The One Blood Rule was a popular social and legal racial classification standard in 20th-century America. It holds that anyone with even a trace of Black blood is identified as Black, not white. This article borrows this concept to illustrate the strict, absolute, no-exception criteria for AI-generated content. Any content containing AI-generated material is required to be labeled, with no exemptions.

first, whether recipients are likely to rely on the content's non-synthetic, authentic, or professionally endorsed nature; second, whether AI-generated information is sufficient to influence judgments about authenticity, responsible parties, or the potential for manipulation. Third, whether disclosure requirements impose fewer constraints on freedom of expression, innovation, and user interests than alternative regulatory approaches. Mandatory identification is warranted only when a substantive connection exists between the source information and the particular risk involved.

**4.3 Refutation of Positive Regulation Theory: Legislative Placebo and the Erosion of Rule of Law Authority**

Another argument is the "positive regulation theory," which recognizes that laws can have technical restrictions or be difficult to enforce, but that it is better to have laws than no laws. It encourages proactive legislative responses to technology's effects on society, even if only symbolic. This view is satisfying to policymakers and the public because it provides a sense of psychosocial certainty through administrative measures. We argue that this approach masks an unwillingness to undertake substantive legislative analysis.

The legislative tendency toward "regulation for the sake of regulation" poses profound risks to the rule of law. Such measures not only fail to solve practical problems but also have significant negative external effects. Technically unenforceable and logically inconsistent laws lead to ineffective regulation. The law's deterrent effect is undermined when the public sees that so much AI-generated content is circulating and that the police are not doing anything about it. This engenders a widespread normalization of non-compliance and resignation toward investigative challenges, which, in turn, erodes public trust in legal authority and leads to a general disregard for compliance.

We argue that symbolic legislation cannot be considered governance. The government's inaction is a problem, and when the government tries to develop only superficial management policies, inevitable governance problems that need to be addressed with significant resource investment, for example, employment transformation in the AI era, algorithmic ethics education, and regulation of data monopolies, might be deferred. Such laws are not only useless but also harmful. It wastes government resources and undermines the rule of law to cater to short-term popular opinion. Over time, it diminishes public trust in the governance system for emerging technologies. Consequently, without a robust legal foundation and empirical evidence, it is preferable to adopt a cautious approach and concentrate oversight on substantive accountability for specific harmful outcomes, including fraud, infringement, and false communication, rather than becoming preoccupied with the formalistic distinction between humans and machines.

**4.4 Summary**

The underlying cause of these three propositions is a lack of up-to-date cognitive models among regulators in post-industrial times. The values and norms in the agricultural society and the early phase of industrialization were relatively simple, allowing for the development of legal norms and rules from social experience and common sense. However, technological advancement rapidly outpaces the cognitive frameworks of both the public and regulators, rendering traditional oversight paradigms obsolete. In contemporary society, the rapid proliferation of complex technologies often causes regulators—limited by their technical expertise—to make flawed policy decisions based on outdated empirical assumptions. The absence of technical expertise during the legislative process frequently yields flawed regulations that are structurally illogical and incapable of achieving meaningful social governance. These errors not only raise compliance costs but also can impede technological advances and contribute to significant social development challenges. Regulators should be mindful that what can be done by legislation need not always be done by legislation, and what might be done by legislation is not necessarily feasible.

Technological philosophy views human creation as a technical embodiment. Writing requires the use of the pen, ink,

and paper, while computing requires the use of the abacus and microchips. In the cyber context in which contemporary content is produced, human cognition and technical tools interact. If writing need not be done by hand, it would be an ontological mistake to exclude AI, an advanced cognitive outsourcing tool, from the creative process based on a binary opposition. This trend of regulatory distinction between human and machine reveals an identity crisis and anxiety about the emergence of post-human creative agents. Labeling machines reiterate the subjectivity given to humans.

## 5. Conclusion: Moving Beyond the Regulatory Placebo

In the face of these challenges and the rapid advancement of technology, the current state of generative AI labeling is prompting regulatory interventions across countries, driven by concerns about information authenticity, accountability, and public trust. This seems to be a compulsory system that can lead to neglect of regulatory enforcement, to over-punitive treatment within legal frameworks, and to unforeseen social effects: accidental harm and technical discrimination.

Given the realities of technological decentralization and cross-border data flows, establishing a compulsory identification system encounters significant jurisdictional and technological enforcement challenges. The prevailing focus on AI-generated labels demonstrates path dependence among regulators as they confront the opacity of advanced technologies. Applying product quality inspection frameworks from the industrial era to regulate AI generation in the information age represents a fundamental misalignment with contemporary technological contexts.

In summary, these policies seek to enforce a strict separation between human creation and AI-generated content through administrative authority, aiming to address social and technological anxieties. While such measures may temporarily placate public anxieties about emerging technologies, they ultimately prove futile and detrimental to the long-term evolution of the rule of law, as they fail to meaningfully engage with the underlying technical realities.

**Author Contributions**

**CHEN Jingyi:** Writing – original draft, Writing – review & editing

**BU Chaofan:** Investigation, Writing – original draft

**YAN Shibo:** Investigation, Writing – review & editing

**LI Xuesong:** Conceptualization, Funding, Investigation, Project administration, Supervision, Writing – original draft, Writing – review & editing

**Acknowledgments**

We would like to thank Prof. OTA Shozo, Célia Filipa FERREIRA MATIAS, Rostam J. NEUWIRTH, FU Liqing, YOSHIMURA Koji, LI Peiwen and MATSUBARA Yoshihiro for their advice. We also thank our colleagues LIU Zhicong, CHEN Sibo, HU Tianhao, IWASE Yuta, MIYAUCHI Koji, FENG Haoyang, WEI Jinrong, WENG Yukai, WU Jinchao, LIN Zhengxiong, LI Kexin, XU Xiaoben, ZHU Xiequn and GUO Danyang for their suggestions.

**Funding**

This work was supported by Meiji University (Grant No. 4122242013).

**Data Availability**

No data was used to support the arguments in this article.

**Declarations**

Conflict of interest: The authors declare no competing interests.

**Reference**

Abdelaziz, D. K. A. (2025). Criminal liability for the misuse and crimes committed by AI: A comparative analysis of legislation and international conventions. Journal of Infrastructure, Policy and Development, 9(1), 10722.

Artificial Intelligence Act. (2024). The Act. https://artificialintelligenceact.eu/

Chen, C., & Shu, K. (2024). Can LLM-generated misinformation be detected? In International Conference on Learning Representations 2024.

Chen, R., et al. (2024). De-mark: Watermark removal in large language models. arXiv preprint arXiv:2410.13808.

Cyberspace Administration of China. (2025, March 14). Notice on seeking public comments on the Administrative Measures for Generative Artificial Intelligence Services. https://www.cac.gov.cn/2025-03/14/c_1743654684782215.htm

Andrew Simmerman (2026, January 10). New California AI laws that matter to you in 2026. EdSource. https://edsource.org/2026/ai-education-policy-california-2025/748653

Beijing DHH Law Firm. (2026, February 3). Analysis of data compliance and copyright protection paths for generative AI. https://www.deheheng.com/content/35793.html

Deshpande, A., & Kanti, V. B. (2025). Insecure AI image watermarking—Is it really damaging the future? In Proceedings of the 9th International Conference on Computer Science and Artificial Intelligence.

El Ali, A., et al. (2024). Transparent AI disclosure obligations: Who, what, when, where, why, how. In Extended Abstracts of the CHI Conference on Human Factors in Computing Systems.

Ghaffary, S. (2023). What will stop AI from flooding the internet with fake images? Vox. https://www.vox.com/technology/23746060/ai-generative-fake-images-photoshop-google-microsoft-adobe

Grinbaum, A., & Adomaitis, L. (2022). The ethical need for watermarks in machine-generated language. arXiv preprint arXiv:2209.03118.

Kirchenbauer, J., Geiping, J., Wen, Y., Katz, J., Miers, I., & Goldstein, T. (2023). A watermark for large language models. In International Conference on Machine Learning.

KeguanJP Editorial Department. (2024, January 25). Japan's MIC and others announce draft AI Governance Guidelines, to be continuously revised. KeguanJP. https://www.keguanjp.com/kgjp_keji/kgjp_kj_ict/pt20240125000003.html

Leuprecht, C., et al. (2021). Patterns in nascent, ascendant and mature border security: Regional comparisons in transgovernmental coordination, cooperation, and collaboration. Commonwealth & Comparative Politics, 59(3), 349–375.

Liu, A., et al. (2024). A survey of text watermarking in the era of large language models. ACM Computing Surveys, 57(2), 1–36.

Ministry of Economy, Trade and Industry. (2024, April 19). Establishment of the AI Governance Guidelines Ver. 1.1. https://www.meti.go.jp/press/2024/04/20240419004/20240419004.html

Ministry of Science and ICT. (2024). Overview of the AI Basic Act. https://aibasicact.kr/

Nah, F. H., et al. (2023). Generative AI and ChatGPT: Applications, challenges, and AI-human collaboration. Journal of Information Technology Case and Application Research, 25(3), 277–304.

Omaar, H. (2024, November 4). California's AI transparency law is a misstep other states should avoid. Center for Data Innovation. https://datainnovation.org/2024/11/californias-ai-transparency-law-is-a-misstep-other-states-should/

Raj, M., Berg, J. M., & Seamans, R. (2026). The artificial intelligence disclosure penalty: Humans persistently devalue AI-generated creative writing. Journal of Experimental Psychology: General.

Reuters. (2023, July 21). OpenAI, Google, others pledge to watermark AI content for safety, White House says. https://www.reuters.com/technology/openai-google-others-pledge-watermark-ai-content-safety-white-house-2023-07-21/

Rijsbosch, B., van Dijck, G., & Kollnig, K. (2025). Adoption of watermarking measures for AI-generated content and implications under the EU AI Act. arXiv preprint arXiv:2503.18156.

Sheng, C., & Hong, H. Y. (2023). Countering, blocking and referring—The US long-arm jurisdiction and China's countermeasures. Journal of WTO & China, 13, 41.

Simard, M. (2024). Position paper: Should machine translation be labelled as AI-generated content? In Proceedings of the 16th Conference of the Association for Machine Translation in the Americas (Volume 1: Research Track).

Stuurman, K., & Lachaud, E. (2022). Regulating AI. A label to complete the proposed Act on Artificial Intelligence. Computer Law & Security Review, 44, 105657.

Susen, S. (2022). Reflections on the (post-) human condition: Towards new forms of engagement with the world? Social Epistemology, 36(1), 63–94.

Wittenberg, C., et al. (2024). Labeling AI-generated content: Promises, perils, and future directions.

Zheng, Y., et al. (2025). A review on edge large language models: Design, execution, and applications. ACM Computing Surveys, 57(8), 1–35.

Zyda, M. (2024). Large language models and generative AI, oh my! Computer, 57(3), 127–132.

California State Legislature. (2025). AI Transparency Act (22757.1(b)).